\documentstyle[epsfig,cite]{mn2e}
\begin{document}

\title[Three episodes of jet activity in B0925+420]
      {Three episodes of jet activity in the FRII radio galaxy B0925+420}
\author[Brocksopp et al.]
    {C.~Brocksopp$^1$\thanks{email: cb4@mssl.ucl.ac.uk}, C.R.Kaiser$^2$, A.P. Schoenmakers$^3$, A.G. de Bruyn$^3$\\
$^1$Mullard Space Science Laboratory, University College London, Holmbury St. Mary, Dorking, Surrey RH5 6NT, UK\\
$^2$School of Physics and Astronomy, University of Southampton, Southampton, Hants. SO17 1BJ, UK\\
$^3$Stichting ASTRON, Postbus 2, 7990 AA Dwingeloo, The Netherlands\\
}
\date{Accepted ??. Received ??}
\pagerange{\pageref{firstpage}--\pageref{lastpage}}
\pubyear{??}
\maketitle
\begin{abstract}
We present Very Large Array images of a ``Double-Double Radio Galaxy'', a class of objects in which two pairs of lobes are aligned either side of the nucleus. In this object, B0925+420, we discover a third pair of lobes, close to the core and again in alignment with the other lobes. This first-known ``Triple-Double'' object strongly increases the likelihood that these lobes represent mutiple episodes of jet activity, as opposed to knots in an underlying jet. We model the lobes in terms of their dynamical evolution. We find that the inner pair of lobes is consistent with the outer pair having been displaced buoyantly by the ambient medium. The middle pair of lobes is more problematic -- to the extent where an alternative model interpreting the middle and inner ``lobes'' as additional bow shocks within the outer lobes may be more appropriate -- and we discuss the implications of this on our understanding of the density of the ambient medium.

\end{abstract}

%*************************************************************************************
\section{Introduction}

The behaviour of Fanaroff-Riley II (FRII; Fanaroff \& Riley 1974) radio galaxies is typified by the ejection of highly collimated, laminar jets from a central core. These jets travel supersonically away from the core and are decelerated by the ram-pressure of the intergalactic medium (IGM). This deceleration results in a bow shock being driven through the IGM, inflating large ``cocoons'' of low-density material, part of which we observe as radio-lobes. Finally the energy is lost in bright hotspots of shocked emission. Although the intrinsic evolution of these powerful radio sources is now largely understood (e.g. Kaiser \& Alexander 1997, hereafter KA; Blundell, Rawlings \& Willott 1999), there are still several unanswered fundamental questions relating to the duty-cycle of the active galactic nucleus (AGN) and the contents of the lobes.

\begin{table*}
\caption{Angular resolutions, position angles (PA) and 1$\sigma$ noise levels in each of the new VLA images.}
\begin{tabular}{lcccc}
\hline
\hline
Frequency   &A-array Resolution (PA)  &A-array Sensitivity     &B-array Resolution (PA)    &B-array Sensitivity\\
(GHz)       & (\arcsec) ($^{\circ}$)           &$\mbox{mJy\,beam}^{-1}$ & (\arcsec) ($^{\circ}$)             &$\mbox{mJy\,beam}^{-1}$\\
\hline
1.4$^*$ & --                             &     --                 & $10 \times 10 $\,(0)             & $3.21\times10^{-1}$\\
1.4     & $1.7\times 1.5$\,(5.1)        & $3.15\times10^{-2}$    & $6.4\times 5.0$\,(68.2)           & $1.70\times10^{-1}$\\
4.9     & $0.6\times 0.5$\,($-28.7$)    & $4.28\times10^{-2}$    & $1.7\times 1.5$\,($-68.8$)        & $2.72\times10^{-2}$\\
8.4     & $0.3\times 0.2$\,($-23.5$)    & $3.20\times10^{-2}$    & $0.9\times 0.8$\,($-68.5$)        & $2.22\times10^{-2}$\\
\hline
\end{tabular}
\begin{tabular}{l}
$^*$The new B-array VLA data, combined with the FIRST data and convolved with a 10\arcsec\, beam as displayed on the left-hand side of Fig.~\ref{images}.
\end{tabular}
\label{resolution}
\end{table*}

\begin{figure*}
\begin{center}
\leavevmode
\epsfig{file=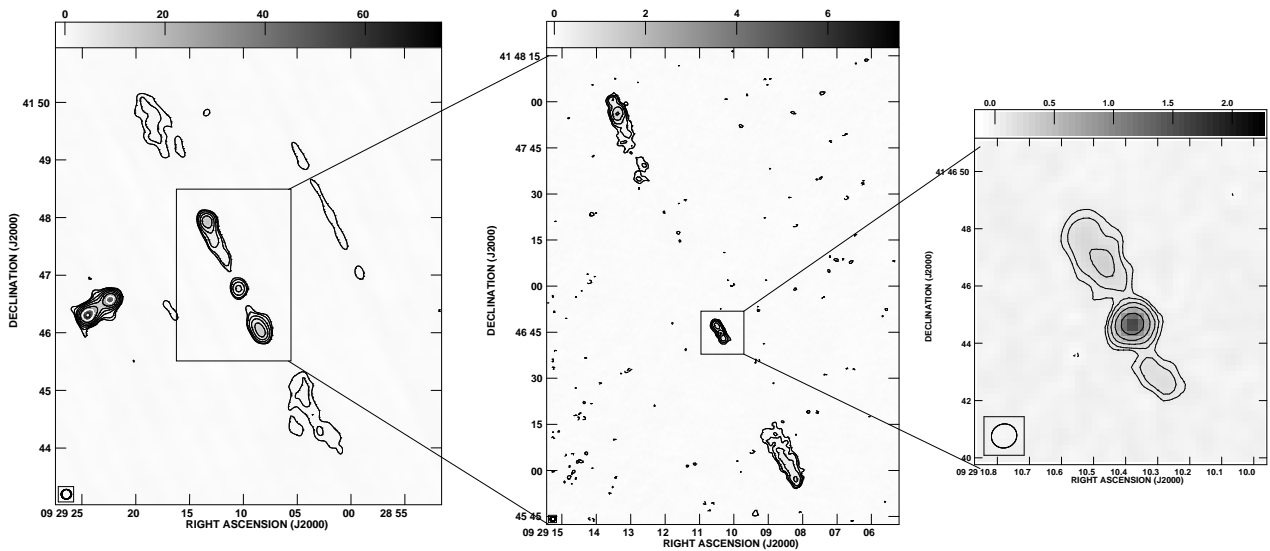, width=17cm}
\caption{VLA images of B0925+420 showing the three pairs of lobes. All contours are at -3, 3, 6, 12, 24, 48$\times \sigma$, with the exception of the left-hand plot which has an additional $2\sigma$ contour. LHS: the 2001 VLA 1.4-GHz B-array data, combined with additional 1.4-GHz B-array data from the 1994 FIRST survey (Becker, White \& Helfand 1995) and finally reconvolved with a 10$\arcsec$ beam in order to detect the weak outer lobes at a 3$\sigma$ level ($\sigma=3.21\times10^{-1}\,\mbox{mJy\,beam}^{-1}$). Centre: 2001 VLA 4.9-GHz B-array image, clearly resolving the middle and inner lobes but resolving out all emission from the outer lobes ($\sigma=2.72\times10^{-2}\,\mbox{mJy\,beam}^{-1}$; angular resolution $=1.7\times 1.5\arcsec$). RHS: 2001 VLA 8.4-GHz B-array image showing a close-up of the inner lobes ($\sigma=2.22\times10^{-2}\,\mbox{mJy\,beam}^{-1}$; angular resolution $=0.9\times 0.8\arcsec$).}
\label{images}
\end{center}
\end{figure*}

\begin{figure*}
\begin{center}
\leavevmode
\epsfig{file=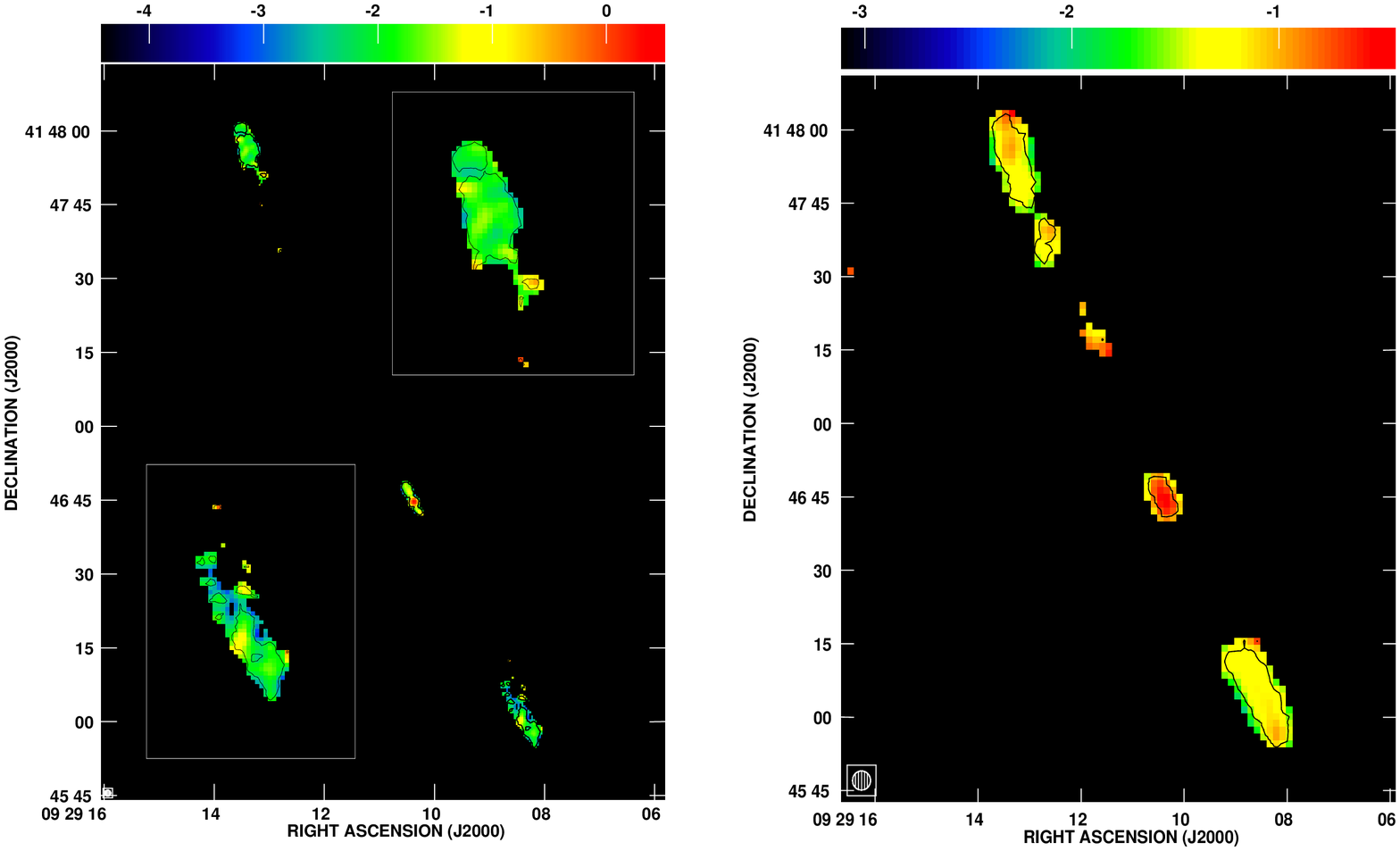, width=17cm}
\caption{Spectral index maps created from the 2001 VLA data; we used only the B-array images so as to ensure that no emission was resolved out at the higher frequencies. In each case, the higher-resolution image was reconvolved with the beam of the lower-resolution image. For comparison with the spatial scale of the original high-resolution image, we overplot each map with the 3$\sigma$ Stokes I flux density as measured in the unconvolved image. LHS: 4.9-GHz/8.4-GHz with angular resolution $=1.7\times 1.5\arcsec$. Inset plots show the northern and southern middle lobes more clearly and we note that the regions of unexpectedly steep spectral index correspond to those regions outside the 8.4-GHz contour. RHS: 1.4-GHz/4.9-GHz with angular resolution $=6.4\times 5.0\arcsec$. The core appears to have a relatively flat spectrum and the lobes a steeper spectrum with $\alpha$ in the range $-1$ to $-2$; the spectrum of the middle lobes appears steeper than that of the inner lobes. }
\label{alpha}
\end{center}
\end{figure*}

``Double-Double Radio Galaxies'' (DDRG) are FRIIs with a second pair of lobes; all four lobes appear to emanate from a common nucleus and are strongly coaligned to within $10^{\circ}$ of each other (although see also Saikia, Konar \& Kulkarni (2006), who identify other DDRG whose lobes are misaligned). When viewed in isolation each of these lobes has a comparable morphology with that of a ``normal'' FRII galaxy. A sample of seven such objects has been identified, studied and modelled by Schoenmakers et al. (2000a, b), Kaiser, Schoenmakers \& R\"ottgering (2000) and Konar et al. (2006); these seven objects are B0925+420, B1240+389, B1450+333, B1834+620, 3C219, 3C445 and 4C26.35 (although in retrospect the inner structures of 3C219 are very narrow and may instead be ``knots'' in the jet rather than new lobes). Saikia et al. (2006) studied a larger sample of DDRG, bringing the total to twelve. The additional pair of radio lobes has been interpreted as evidence for a second phase of jet activity, the inner pair of lobes forming within the low-density cocoon of the original jet. With this unusual opportunity to obtain information about multiple episodes of activity, this subset of the FRII class has the potential to give us rare insight into the conditions of the surrounding medium before and after the launching of the jet, as well as duty-cycle timescales.

There are still uncertainties regarding certain aspects of these objects. Are the inner structures truly separate radio sources with no physical connection to the outer lobes? Are the ages and ambient densities of the inner lobes -- which may be the density of the cocoon inflated by the original jet -- atypically low, i.e. much lower than these values are for objects of similar size as the inner radio lobe? Are there any remnants of jets and hotspots in the outer lobes? Is the density much lower than in the ``normal'' galactic environment, which would strongly suggest that the inner structure is indeed embedded in the ``hole'' excavated by the outer lobes? Or is the density higher which would imply very efficient replacement of the inner lobes by the heavier external medium. Either result would be extremely important because of the constraints placed on the physics.

In this paper we study one such DDRG, B0925+420 (=J0929+4146), which provides particularly strong evidence for multiple episodes of activity. We present the observations in Section 2, analysis of the images in Section 3 and details of the polarisation properties in Section 4. In Section 5 we present the results of analytical modelling of each of the radio lobes. We make comparisons with other radio transients in Section 6 and finally draw our conclusions in Section 7.

\section{Observations}

Radio observations of B0925+420 were obtained using the VLA on 2000 October 20--21, when the array was in the highest resolution A-configuration, and 2001 April 2, when the array was in the B-configuration. Data were obtained at three frequencies -- 1.4-GHz (L-band), 4.9-GHz (C-band) and 8.4-GHz (X-band) -- in each array; IFs at two adjacent frequencies were used in each continuum observation. In order to reduce potential bandwidth-smearing in the more extended images, all A-array observations and the 1.4-GHz B-array observations were obtained in the PA/PB spectral line observing mode. This gave us 8 channels of 3.125 MHz each, resulting in a 25-MHz bandwidth instead of the more usual, continuum bandwidth totalling 50 MHz. Full polarisation information was also recorded. 3C\,286 (J1331+305) was used as the flux and polarisation calibrator while JVAS~0920+446 was used for phase calibration and to ensure parallactic angle coverage.

Additional images were obtained from the Faint Images of the Radio Sky at 20-cm (FIRST; Becker et al. 1995), NRAO VLA Sky Survey (NVSS; Condon et al. 1998) and Westerbork Northern Sky Survey (WENSS; Rengelink et al. 1997) surveys at 1.4-GHz, 1.4-GHz and 0.33-GHz respectively.

The 2000/2001 VLA data were reduced using standard flagging, calibration, cleaning and imaging routines within {\sc aips}. The flux densities of 3C\,286 were obtained using the formulae of Baars et al. (1977) but with the revised coefficients of Rick Perley, as is the default option in the {\sc setjy} routine. Additional analysis routines within {\sc aips} were used to convolve the images with different beam-sizes, combine the new 1.4-GHz data with the FIRST survey image, measure lobe parameters and create spectral index maps. We list the angular resolutions and sensitivities of the images at all three frequencies and in both array-configurations in Table~\ref{resolution}.

\begin{table*}
\caption{Dimensions and flux densities ($S_{\nu}$) of the lobes obtained from the images. Typically the lower-resolution images were used so as to reduce the likelihood of resolving out any flux. Errors are to within up to $\sim 5\arcsec$ and $\sim 1$ mJy. Lobe-lengths have been measured from the leading edge to the centre of the core; widths have been measured at their widest points in the non-convolved images. The last two columns list the spectral indices in the 1.4-GHz / 4.9-GHz and 4.9-GHz / 8.4-GHz bands. These values are calculated from the flux densities and are broadly comparable with the spectral index maps shown in Fig.~\ref{alpha}; we note that (i) the calculated spectral index for the southern middle lobe is artificially steep due to some flux being resolved out at 8.4-GHz, and (ii) the calculated spectral index of $-1.22$ for the core is artificially steep due to contamination by emission from the inner lobes at low resolution.}
\begin{tabular}{lcccccccc}
\hline
\hline
Lobe         & Length   & Width    & $S_{0.32\,\mbox{MHz}}$      &  $S_{1.4\,\mbox{GHz}}$  & $S_{4.9\,\mbox{GHz}}$      & $S_{8.4\,\mbox{GHz}}$     &$\alpha$           & $\alpha$\\
             & (arcsec) & (arcsec) & (mJy) & (mJy)&  (mJy)  & (mJy) & (1.4-GHz/4.9-GHz) & (4.9-GHz/8.4-GHz)\\
\hline
North, outer & 227     &   38      & 88                & 23              & --                &  --              &    --        &      --       \\
South, outer & 193     &   42      & 84                & 24              & --                &  --              &    --        &      --     \\
North, middle& 83      &   4       & --                & 32              & 10                &  4               & $-0.93$      &      $-1.70$       \\
South, middle& 57      &   4       & --                & 29              & 10                &  2               & $-0.85$      &      $-2.99$       \\
North, inner & 4       &   2       & --                & --              & 1.4               &  1.0             &    --        &      $-0.62$       \\
South, inner & 3       &   1       & --                & --              & 0.5               &  0.3             &    --        &      $-0.95$       \\
Core         & --      &   --      & --                & 6               & 1.3               &  1.4             & $-1.22$      &      $0.14$       \\ 
\hline
\end{tabular}
\label{data}
\end{table*}

\section{Results -- Images}

B0925+420 is detected in the 2000/2001 VLA data at each of the six frequency/resolution combinations. While much of the extended emission is resolved out in the higher-resolution A-array images, the remaining images clearly show a series of four well-aligned FRII-like lobes. Closer inspection reveals that the outer lobes detected in the FIRST image (Schoenmakers et al. 2000a) are resolved out in the more recent images and that the core has been resolved at high frequency into an additional, third, pair of lobes.

We have obtained the 1.4-GHz FIRST image and combined the data with the more recent 1.4-GHz data. Finally, by convolving this resultant image with a larger beam (10$\arcsec$) we are able to detect the outer lobes at a 3$\sigma$ level. This reconvolved image is shown on the left-hand side of Fig.~\ref{images}. We note that the morphology of the outer lobes is consistent with that shown in Schoenmakers et al. (2000a); it is simply the choice of contour levels which leads to this apparent inconsistency. The second pair of lobes is resolved clearly while the core remains unresolved. The source to the left of the image is unrelated to B0925+420. Comparison with the NVSS image suggests that no flux is resolved out of the middle lobes in the new VLA B-array images at 1.4-GHz.

The middle of Fig.~\ref{images} is the 4.9-GHz B-array image. The outer lobes are resolved out but the middle pair of lobes are shown clearly. The core is now resolved into a smaller pair of lobes. Finally, the right-hand side of Fig.~\ref{images} shows the 8.4-GHz B-array image; even the inner lobes are resolved out and/or undetected in the A-array image. A bright core is still detected, suggesting the possibility that either these lobes are still being fed by jets or that the core may yet be further resolved into more components.

\begin{figure}
\begin{center}
\leavevmode
\epsfig{file=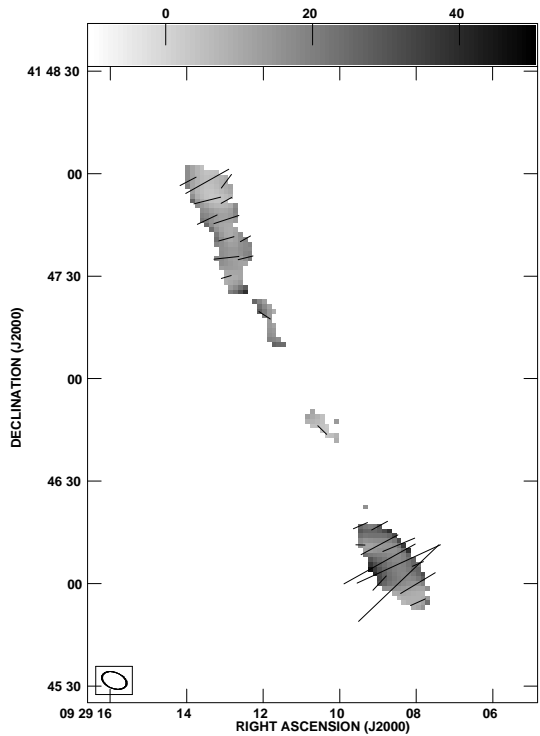, width=7cm}
\caption{Fractional linear polarisation (FP) image at 1.4-GHz using the 2001 VLA:B-array data only; FP is represented by the greyscale. The vectors indicate the direction of the electric field, following correction for RM, and their length indicates the linear polarisation ($1\arcsec = 4.46\times10^{-5}\mbox{Jy\,beam}^{-1}$). The core and inner lobes are weakly polarised; the middle lobes are polarised at the $\sim10-20\%$ level, higher in the southern lobe and around the lobe edges. The position angle of the vectors indicates that the magnetic field in the middle lobes is aligned with the long axis of the lobe, although there appears to be some rotation in the inner lobes.}
\label{fp}
\end{center}
\end{figure}

Following the arguments outlined above for the ``double-double radio galaxies'' providing the best-known evidence for the duty cycle of jets in radio galaxies, {\bf we propose that B0925+420 is therefore the first known radio galaxy which has been observed to display three episodes of jet activity}. This ``triple-double'' radio galaxy provides us with the strongest evidence yet that these objects are indeed potential laboratories for the study of the duty cycle of AGN jets, as opposed to a sequence of bright knots in an underlying jet structure. We discuss this suggestion quantitatively in Section 5.

Finally we have created spectral index maps from the 1.4-GHz / 4.9-GHz and 4.9-GHz / 8.4-GHz images; we define the spectral index, $\alpha$, in terms of $S_{\nu}\propto\nu^{\alpha}$, where $S_{\nu}$ is the flux density at frequency $\nu$. In order to reduce the chances of flux being resolved out, only the B array images were used. In each case, the higher-frequency image was reconvolved with the beam of the lower-frequency image and the resultant images presented in Fig.~\ref{alpha}. The core shows a very flat spectrum with $\alpha$ close to zero. The lobes show steeper spectra of $-1$ to $-2$, with the middle lobes appearing to have steeper spectra than the inner lobes. The spectral index of the southern middle lobe, in particular, appears artificially steep. If we assume a value of $\alpha_{4.9-GHz/8.4-GHz}=-2$, it suggests that the expected flux density at 8.4-GHz is $\sim 3-4$ mJy and so $\sim2$ mJy of flux has been resolved out; similarly, assuming a spectral index of -1 suggests that $\sim5$ mJy has been resolved out. These estimates require that no flux is resolved out at 4.9-GHz, a reasonable assumption given the uniformity of the southern middle lobe in the $\alpha_{1.4-GHz/4.9-GHz}$ map.

While the bright core suggests the possibility that the inner lobes are still receiving energy via jets (as also found in other objects, e.g. J0041+3224; Saikia et al. 2006), the absence of hotspots in the outer and possibly the middle lobes suggests that these structures are no longer supplied with energy. Such a statement is supported by the steep spectral indices presented in Fig.~\ref{alpha} (middle lobes) and the table of observed parameters (Table~\ref{data}), which includes additional data from the Westerbork array and thus provides a second frequency at which the outer lobes were detected. Theoretical modelling in Section 5 adds further support to these observations. We note that we cannot categorically rule out the presence of hotspots in the middle lobes -- some 3CR objects have recessed hotspots, and structures of sizes comparable with the resolution limit have been detected in other objects (e.g. Jeyakumar \& Saikia 2000) -- but, with no firm evidence to the contrary, we assume that they are not present in the middle or outer lobes of B0925+420.

\section{Results -- Polarisation}
The middle and inner lobes of B0925+042 were also detected in the linear polarisation (LP) images made from the 2001 VLA data in B array. We do not have polarisation calibration available for the FIRST data and so cannot extend this analysis to the outer lobes. The LP flux density at 4.9- and 8.4-GHz was low and/or the structure partially resolved out but the structure is clear at 1.4 GHz and, where structure was visible, the fractional polarisation (FP) did not appear to vary significantly with frequency. A rotation measure (RM) image was created, using all three frequencies. The lowest possible RM values were in the range 10--15 rad\,m$^{-2}$, which is consistent with the Galactic medium being the only Faraday screen between us and the source; such a conclusion was also reached for B1834+620 (Schoenmakers et al. 2000b) and any intrinsic contribution to the RM in J1453+3308 was deemed to be small (Konar et al. 2006). We show the 1.4-GHz FP image in Fig.~\ref{fp}; electric field vectors are over-plotted and include the correction ($28^{\circ}$) due to Faraday rotation.

The middle lobes are clearly polarised to 10--20\%, higher in the southern lobe and towards the edges; such a FP is comparable with that measured for J1453+3308 (Konar et al. 2006). The core and/or inner lobes of B0925+480 are weakly polarised. The direction of the electric field vectors appears to change between the two pairs of lobes -- at the centre, the magnetic field is apparently aligned quasi-perpendicularly to the position angle of the radio galaxy and direction of jet motion. The magnetic field in the middle lobes has become aligned comparably with the radio galaxy. We note that this effect in the core/inner lobes is ambiguous since they are not fully resolved and we may be looking at the combined effects of an unpolarised core and polarised inner lobes. This structure of the magnetic field in B0925+480 seems less complex than in J1453+3308, in which the vectors rotated at the edge of the lobe (Konar et al. 2006); however, as discussed in Sections 5, we highly suspect that the outer lobes of J1453+3308 exist in a very different environment from the middle lobes of B0925+480 and so we are not comparing like with like.

The low RM and high FP of B0925+480 suggest that there is no evidence for depolarisation in the middle lobes, hinting that the environment is not very dense and/or clumpy. With their lower levels of LP, the inner lobes could still be dense and clumpy. The modelling in the next section provides further clues as to the nature of the environment of each pair of lobes, confirming the results of the polarisation observations. The alignment of the magnetic field with the edge of the middle lobes may be indicative of compression of the field by the lobe expansion. Surprisingly, given the absence of evidence for a current energy supply, this would argue for continuing expansion of the middle lobes.

\begin{table*}
\caption{Model parameters used in the fitting process.}
\begin{tabular}{clcc}
\hline
\hline
Parameter & Description & Nature of & Value if set\\
          &             &parameter  & {\em ab initio}\\
\hline
$L'$ & Projected length of lobe & observed & --\\
$P_\nu$ & Luminosity density of lobe at observing frequency $\nu$ & observed & --\\
$R'$ & Aspect ratio of lobe & observed & --\\
$z$ & Source redshift & observed & 0.365\\
$\Gamma _{\rm c}$ & Adiabatic index of lobe material & set & $4/3$\\
$\Gamma_{\rm B}$ & Adiabatic index of magnetic field `fluid' & set & $4/3$\\
$\Gamma_{\rm x}$ & Adiabatic index of external medium & set & $5/3$\\
$\alpha$ & Viewing angle of the jet axis to our line of sight & set & $\pi / 2$\\
$k$ & Energy density of non-radiating particles in the lobe & set & 0\\
$\delta$ & Exponent of initial power-law energy density of relativistic electrons & set & $2\rightarrow 2.5$\\
$\gamma_{\rm min}$ & Minimum Lorentz factor of relativistic electrons at time of injection into the lobe & set & 1\\
$\gamma_{\rm max}$ & Maximum Lorentz factor at same time & set & $10^8$\\
$\beta$&Exponent for the power-law density distribution & set & 1.5\\
$a_0$ & Core radius of density distribution of external medium & set & 2\,kpc\\
$\rho_0$ & Density of external medium at a distance $a_0$ from the AGN & fitted & --\\
$Q$ & Jet power & fitted & --\\
$t$ & Time elapsed since jet flow started & fitted & --\\
$t_{\rm s}$ & Time at which jet flow is interrupted & fitted & --\\
\hline
\end{tabular}
\label{modpara}
\end{table*}

\begin{table*}
\caption{Input parameters for the model. Angular sizes and flux densities were converted into luminosity densities and lengths from the values in Table~\ref{data} and using a value of 0.365 for the redshift, $z$. A WMAP cosmology was assumed, with $H_0=71\mbox{km\,s}^{-1}\,\mbox{Mpc}^{-1}$, $\Omega _{\mbox m}=0.27$, $\Omega _{\Lambda}=0.73$ (Spergel et al. 2003). We note that the luminosity at 1.4-GHz for the outer lobes has been derived from the flux density quoted in Schoenmakers et al. (2000a); it appears that some flux was resolved out of the images presented here.}
\begin{tabular}{lcccccc}
\hline
\hline
Lobe & Length & Aspect Ratio & Freq. 1 & Lum. Density 1          & Freq. 2 & Lum. Density 2\\
     & (kpc)  &              &  (GHz)  & ($10^{24}$\,W\,Hz$^{-1}$) & (GHz)     & ($10^{24}$\,W\,Hz$^{-1})$\\
     & $L$    & $R$          & $\nu_1$ & $P_{\nu_1}$               &  $\nu_2$   & $P_{\nu_2}$\\
\hline
North, outer & 1150 & 6   & 0.32& 25    & 1.4& 10   \\
South, outer & 972  & 4.6 & 0.32& 24    & 1.4& 11   \\
North, middle & 419 & 21  & 1.4 & 15    & 4.8& 4.4  \\
South, middle & 287 & 14  & 1.4 & 13    & 4.8& 4.5  \\
North, inner & 20.2 & 2.6 & 4.8 & 0.64  & 8.5& 0.45 \\
South, inner & 15.6 & 2.8 & 4.8 & 0.23  & 8.5& 0.14 \\
\hline
\end{tabular}
\label{obpara}
\end{table*}

\section{Modelling the source-size and luminosity}
\subsection{Brief description of the model}

We use the model of KA for the dynamical evolution of the large-scale structure of extragalactic radio sources with a FRII-type morphology. The model envisages two jets propagating in opposite directions from a central AGN. The ram pressure of the jet material is balanced by the ram pressure of the receding external medium in front of the jet flow. This interaction between the jet and its environment gives rise to a strong shock at the end of the jet and a bow shock in the external gas. After passing through the shock, the jet material inflates a lobe or cocoon surrounding the jet flow. 

The thermal pressure inside the lobe, $p$, balances the pressure of the jet material and thus confines the jet flow. The lobe pressure also drives the sideways expansion of the lobe. In the model of KA the lobes are overpressured with respect to the external medium and so the sideways expansion of the lobes is also confined by the ram pressure of the receding external medium. In this way, the bow shock in front of the jet shock is extended to envelope the entire lobe structure. The density distribution in the external medium is modelled as a power-law centred on the AGN with 

\begin{equation}
\rho = \rho_0 \left( \frac{r}{a_0} \right)^{-\beta},
\end{equation}

\noindent
where $\rho_0$ and $a_0$ are constants and $r$ measures the distance from the AGN. The model then predicts the length of the lobe in the direction of the jet flow, $L$, as a function of the age of the flow, $t$, the energy transport rate of the jet or jet power, $Q$, the exponent for the power-law density distribution, $\beta$, and the `density parameter', $\rho_0 a_0^{\beta}$. Another parameter entering the calculations is the aspect ratio of the lobe, $R$, which we define here as the ratio of $L$ and the width of the lobe, $W$, measured at the widest point along $L$ so that $R=L/W$.

The model of KA was extended by Kaiser, Dennett-Thorpe \& Alexander (1997, hereafter KDA) to allow the prediction of the radio luminosity density of the lobe, $P_{\nu}$, as a function of essentially the same parameters. The KDA model assumes that the radiating relativistic electrons are injected into the lobe at the jet shock with a power-law energy distribution according to $n(E) \, {\rm d}E \propto E^{-\delta} \,{\rm d} E$. The subsequent energy losses of the electrons due to the adiabatic expansion of the lobe and radiative processes, emission of synchrotron radiation and inverse Compton scattering of cosmic microwave background photons, are taken into account self-consistently. 

The middle and (particularly) the outer lobes of B0925+420 show little evidence for the presence of radio hotspots which are usually identified as the impact sites of the jets and the location for the injection of relativistic electrons into the lobes. It is therefore likely that the energy supply of the jets as well as the injection of relativistic electrons into these lobes has ceased some time ago. If we assume that some disturbance of the jet flow close to the AGN is responsible for stopping the jet at a time $t_{\rm s}$ after the onset of jet activity, then the injection of relativistic electrons into the lobe stops at $t_{\rm s} + t_{\rm t}$, where $t_{\rm t}$ is the time it takes the last jet material to travel from the AGN to the end of the jet. Assuming that the jet material travels with a speed close to the speed of light inside the jet, we have $t_{\rm t} = L/c$ where $L$ is a function of $t_{\rm s} + t_{\rm t}$. Using the model of KA we can solve this implicit relation for $t_{\rm t}$ by iteration.

When the energy supply of the jet to the lobe shuts down at time $t_{\rm s}+t_{\rm t}$, the dynamics of the lobe expansion start to change (e.g. Kaiser \& Cotter 2002). However, this new expansion regime takes a few sound crossing times of the lobe to establish itself. For the large sizes of the middle and outer lobes the time for this adjustment is comparable to the total source age. Therefore for simplicity we continue to use the model of KA for the expansion of the lobe. However, when calculating the luminosity density of the lobe we only take into account relativistic electrons injected before the time $t_{\rm s} + t_{\rm t}$. The same approach was used in Kaiser et al. (2000).

\subsection{Fitting process}

We can use the models of KA and KDA to constrain the properties of the jets and the environment of B0925+420 from the observed properties of its lobes. Table~\ref{modpara} summarises the model parameters used in this process. Most of the model parameters we do not discuss in detail here, since they are described in the papers by KA and KDA. {\em Observed\/} parameters are those taken from observations and some of these obviously vary from lobe to lobe. {\em Set\/} parameters describe the hydrodynamic and other properties of the lobe contents and the external medium. They are set to reasonable values and are not varied in the fitting process. {\em Fitted\/} parameters are the free parameters of the model constrained by the observational data. 

We will see below that the very large observed size of the outer lobes of B0925+42 imply very old ages for the jet flows. If the jets and therefore the lobes are oriented at small angles to our line of sight, then the actual lobe length will be even larger. Hence we adopt a viewing angle $\alpha =\pi /2$, i.e. the lobes are in the plane of the sky and the actual lobe length and aspect ratio equal the observed values. The model depends only on the density parameter $\rho _0 a_0^{\beta}$ rather than on $a_0$ and $\rho_0$ separately. For convenience we set $a_0=2$\,kpc, a value appropriate for isolated elliptical galaxies, and $\beta=1.5$ (Fukazawa, Makishima \& Ohashi 2004) and only fit the value of $\rho_0$. In comparing model results we therefore note that a change of $\rho_0$ may also imply a change in $a_0$. Finally, the slope of the power-law energy distribution of the relativistic electrons strongly influences the slope of the observed radio spectrum. We adjust the value of $\delta$ in the range 2 to 2.5 as discussed in the next sub-section.

For each individual lobe we use the lobe length and aspect ratio as well as the luminosity densities measured at two frequencies in the fitting process. Table~\ref{obpara} summarises the model inputs derived from our observations.

\subsection{Model results}

\begin{figure}
\begin{center}
\leavevmode
\epsfig{file=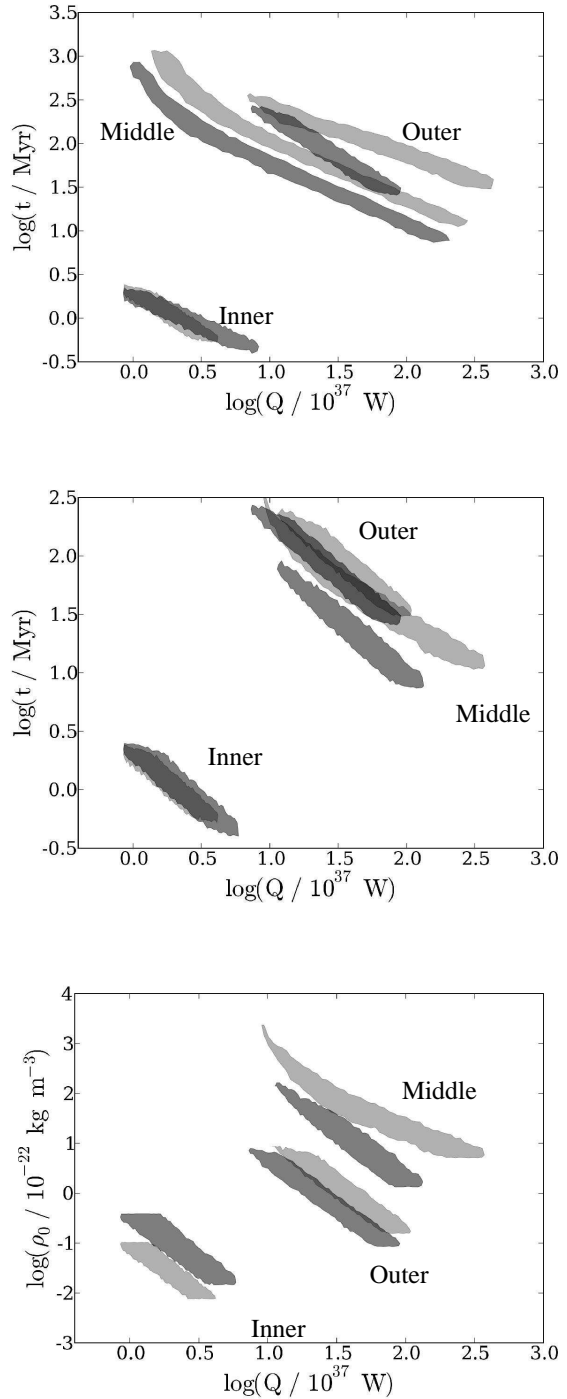, width=8cm}
\caption{Model solutions for all lobes with, in each case, light grey patches representing the northern lobe of each pair and the dark grey patches representing the southern lobes. Top: allowed combinations of jet power and source age for $\delta=2$. Middle: allowed combinations of jet power and source age when we change $\delta$ to 2.5 for the middle lobes and the southern, inner lobe, plus change $\delta$ to 2.2 for the northern, outer lobe. Bottom: allowed combinations of jet power and the density of the external medium for the same values of $\delta$ as in the middle panel.}
\label{model}
\end{center}
\end{figure}

To determine the model parameters $\rho_0$, $Q$, $t$ and $t_{\rm s}$, we randomly choose a large number of combinations of these four parameters and calculate the prediction of the model for the lobe length $L$ and the luminosity densities $P_{\nu}$ at two observing frequencies. A given parameter combination is deemed to be consistent with the observations, if the model results are all within 10\% of the observed values of $L$ and the two values of $P_{\nu}$. Each lobe is fitted individually and the results for the source age $t$ and the jet power $Q$ are shown in the top panel of Fig.~\ref{model} where we have set $\delta =2$ for all lobes.

It is re-assuring that for most possible jet powers the lobe age reflects the lobe size with the outer lobes the oldest and the inner lobes the youngest. We would expect that the jets inflating each pair of lobes, outer, middle and inner, have the same jet power and age on both sides of the AGN. Hence we expect the patches in Fig.~\ref{model} (top) for a given pair to show at least some overlap. While this is the case for the inner and outer lobe pair, the middle pair shows no overlap at all. Also, the area in the $Q$-$t$ plane allowed for both lobes of the outer pair is only small. This may be a result of a wrong choice for the slope of the initial power-law energy spectrum of the relativistic electrons, $\delta$, for some of the lobes. The radio spectra of the middle lobes are steep compared to the spectra of most of the other lobes and may imply a steeper initial energy spectrum. In the interests of exploring the parameter space, in the middle panel of Fig.~\ref{model} we show the result of changing $\delta$ to 2.5 for the middle lobes. We also change $\delta$ to 2.2 for the northern, outer lobe and to 2.5 for the southern, inner lobe. Both of these also show at least slightly steeper radio spectra. This demonstrates the effect of changing $\delta$ on the model results.

The agreement of the model parameters for the two outer lobes is now increased, as is that for the inner lobes. However, the model is unable to reconcile the parameters of the middle lobes, even for values of $\delta$ greater than 2.5. A further problem with the middle pair of lobes is illustrated in the bottom panel of Fig.~\ref{model} which shows the allowed combinations of $Q$ and the density of the external medium, $\rho_0$. For a given jet power the model requires the density of the gas in the environment of the middle lobes to exceed that of the outer lobes. The same result is found for different values of $\delta$. The observed source geometry strongly suggests that the middle lobes are located in a region where the outer lobes have displaced the ambient medium. We argue below that at least some of this region has subsequently been buoyantly refilled with ambient gas, but it is very unlikely that this refilling would lead to a substantial increase in the gas density compared to the undisturbed gas distribution. This suggests that the model described here cannot adequately fit the observed properties of the middle lobe pair.

\begin{figure}
\begin{center}
\leavevmode
\epsfig{file=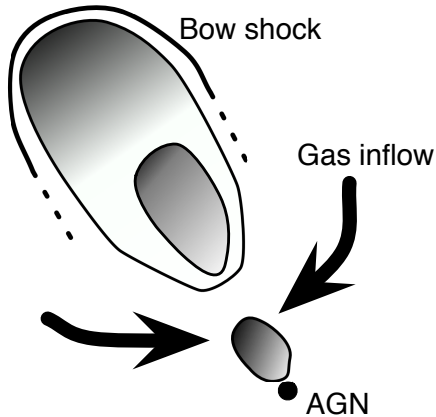, width=6.5cm}
\caption{Schematic diagram illustrating the results of the modelling. The outer lobe, containing the middle lobe, has started to rise buoyantly and is being replaced with the heavier ambient medium. The darker shading represents the part of the lobe which may still be over-pressured with respect to the surrounding medium and so may continue to drive a bow-shock.}
\label{schematic}
\end{center}
\end{figure}

In contrast the inner lobe pair is consistent with a somewhat reduced gas density compared to the situation for the outer lobe pair. Note that asymmetries in the gas distribution around the host galaxy are very likely on scales of several to hundreds of kpc. The absence of any significant overlap of the patches for the inner and outer lobe pairs in Fig.~\ref{model} (bottom) is therefore acceptable. 
Finally, if the disruption of the jet flow to individual lobes is indeed caused by processes close to the AGN, then we would expect the time $t_{\rm s}$ at which the disruption occurs to be the same for both lobes in a given pair. We find that this is the case for the inner and outer lobe pairs for the values of $\delta$ discussed above. 

We can derive another, independent estimate for the power of the currently-active jets. In our model we assume that the inner lobes are currently supplied with energy by active jets from the AGN. These jets also give rise to the radio emission of the unresolved core in our highest resolution radio maps. We measure the flux density of the core as 6, 1.3 and 1.4 mJy at 1.4, 4.9 and 8.4 GHz respectively. The radio spectrum is therefore flat at least at high frequencies. The greater flux density at the lowest observing frequency is probably caused by unresolved flux of the inner lobes contaminating the core flux. The jet power of the currently active jet can be estimated from the flat spectrum core flux as $1.1\times 10^{37}$\,W (K\"ording, Fender \& Migliari 2006). This is consistent with the lower end of the power estimates arising from the observed lobes. 

Our findings can be summarised as ranges of model parameters consistent with the observed properties of the inner and the outer lobe pairs. They are as follows:

\begin{itemize}
\item The power of the jets which inflated the outer lobes was in the range $10^{38}$\,W to $8 \times 10^{38}$\,W while the inner lobe pair implies a jet power of $10^{37}$\,W to $4\times 10^{37}$\,W. A similar jet power of $1.1\times 10^{37}$\,W is estimated from the core emission. The jet power appears to have decreased over time.
\item The density at $a_0$ in the environment of the outer lobes is in the range $8\times 10^{-24}$\,kg\,m$^{-3}$ to $8\times 10^{-22}$\,kg\,m$^{-3}$. For the inner lobes this decreases to $7\times 10^{-25}$\,kg\,m$^{-3}$ up to $4\times10^{-23}$\,kg\,m$^{-3}$.
\item The age of the outer lobes lies in the range 25 to 270\,Myr while that of the inner lobes is 0.4 to 2\,Myr.
\item The jet flow supplying the outer lobes with energy is predicted to have ceased between 4 and 70\,Myr ago. The properties of the inner lobes are consistent with a currently continuing energy supply through active jets.
\end{itemize}

\subsection{Discussion of model results}

The model results for the outer lobe pair are broadly consistent with the properties of other old radio galaxies with an FRII-type morphology. The middle and inner lobes appear to evolve inside the outer lobes and we describe here qualitatively how this may take place (see also Fig.~\ref{schematic}). 
 
While the outer lobe is young and being inflated, it is overpressured with respect to the ambient medium. Thus the lobe drives a bow shock into the ambient medium, compressing and displacing it. Once the jet has stopped supplying energy to the lobe, the lobe gradually comes into pressure-equilibrium with the ambient medium. Those regions of the lobe in pressure equilibrium are no longer surrounded by a bow shock and they represent simply a bubble of light gas (lobe) immersed in the heavier ambient medium at the same pressure. The pressure balance is initially achieved close to the core where the pressure in the ambient medium is highest; the central pinching of the lobe first discussed by Baldwin (1982). Note that the outer regions of the lobe may still be overpressured at this stage and therefore drive a bow shock. The parts of the lobe closest to the centre start to rise outwards due to buoyancy forces and are replaced by the heavier ambient medium sinking towards the core (labeled ``Gas inflow'' in Fig.~\ref{schematic}). 

Buoyant replacement of the outer lobes of B0925+420 sets in when the pressure inside these lobes becomes comparable with the thermal pressure in the lobe environment. We can calculate a maximum pressure of $9\times10^{-15}$\,J\,m$^{-3}$ inside the outer lobes allowed by the model parameters found in the previous sub-section. If we assume isothermal conditions in the lobe environment with a gas temperature of typically $10^7$\,K, then the thermal pressure of the ambient gas exceeds that inside the outer lobes at least out to a distance of 25\,$a_0 \sim 50$\,kpc. The buoyant replacement of the lobe proceeds at the sound speed of the ambient gas, about 400\,km\,s$^{-1}$ for the assumed temperature. It then takes about 40\,Myr to buoyantly replace the outer lobe out to 20\,kpc which is comparable with the size of the inner lobes. The same process would take roughly 200\,Myr to replace the outer lobe out to 100\,kpc. This is still considerably closer to the centre than the end of the middle lobes. Given the age of the outer lobes it is possible that the environment of the inner lobes is created by the buoyant replacement of the outer lobes. However, the same process is unlikely to be fast enough to provide the environment for the middle lobes. 

The middle lobes will initially have to plough their way through the ambient gas that has replaced the outer lobes. They may then reach the lower boundary of the outer lobes and continue to expand inside them. However, the modelling results suggest that (a) the middle lobes have also been replaced by buoyant motions, and (b) the inner lobes are currently expanding in that new ambient medium. The poor fit of our model to the observed properties of the middle lobes suggests that the radio synchrotron emission may not be that of the jet-inflated middle lobes themselves. 

\subsection{Alternative model for the middle and inner lobe pairs}

Once the jets responsible for the middle lobe pair enter the outer lobes, they are immersed in a low density environment. Clark \& Burns (1991) simulated a restarting jet that enters the lobe it inflated in a previous activity phase. They find that the jet propagates almost ballistically through the old lobe and it does not form a significant new lobe inside the old one. However, they find a narrow bow shock driven by the jet into the material of the old lobe. This bow shock compresses the lobe material and may also re-accelerate relativistic particles. The observed middle lobes of B0925+420 are narrow and in the following we interpret them as the bow shocks surrounding the jets within the old lobes. 

We can derive the properties of the magnetised plasma in the outer lobes by assuming minimum energy conditions in these lobes (e.g. Longair 1984). For simplicity we assume that the power-law energy distribution of relativistic electrons, in terms of the electron Lorentz factor, extends from $\gamma =1$ to $\gamma \rightarrow \infty$ with an exponent of 2.5. We also assume a cylindrical geometry for the outer lobes with the emitting region at 326\,MHz extending about three quarters of the way back from the leading edge towards the source centre. We then derive the magnetic field strength for both outer lobes as $B_i = 0.2$\,nT, where $i$ refers to the initial, or pre-shocked plasma. The integrated density of electrons in the assumed energy distribution is then $n_i=0.09$\,m$^{-3}$. We use the same assumptions for the middle lobes, except that the radio emission of these structures observed at 1.4\,GHz only extends to about halfway between the leading edge of each lobe and the source centre. We find a magnetic field strength of $B_s = 1$\,nT for both middle lobes (i.e. the shocked plasma) which implies an electron density of around $n_s=2$\,m$^{-3}$.

If the radio emission of the middle lobes arises from the shock-compressed material in the outer lobes, then $B_s$ and $n_s$ must be related to $B_i$ and $n_i$ by the relativistic shock jump conditions (e.g. Landau \& Lifshitz 1987). We can test this by deriving a value for the shock velocity (in units of $c$, the speed of light) in the rest-frame of the unshocked plasma in the outer lobe from 

\begin{equation}
\beta_{\rm s} = \sqrt{\frac{ \left( p_s - p_i \right) \left( e_s + p_i \right)}{\left( e_s - e_i \right) \left( e_i + p_s\right) }},
\end{equation}

where $p_i, p_s$ are the plasma pressures measured in the respective rest-frames of the initial and shocked plasma respectively. The energy densities $e_i, e_s$ are defined as $e = n \bar{m} c^2 + u$, with $\bar{m}$ the average rest-mass of a gas particle and $u$ the thermal energy density, again measured in the rest-frame of both the initial and the shocked plasma. For simplicity we assume that the outer lobes contain either an ultra-relativistic pair plasma with $u = 3 p$ and $\bar{m} = m_{\rm e}$ or a proton-electron plasma with one cold proton for each relativistic electron with $u = (3/2) p$ and $\bar{m} = m_{\rm p}$.

In both cases the pressure terms are dominated by the magnetic field and the relativistic particles and it then follows from the assumption of minimum energy conditions $u = 7  B^2 / \left( 6 \mu_0 \right)$, with $\mu_0$ the permeability of the vacuum. For the case of a pure pair plasma we find the unphysical result of $\beta_{\rm s} > 1$. Clearly the mass density of a pure pair plasma is too low to be consistent with the shock jump conditions. In other words, a pure plasma in the outer lobes is insufficient to stop the jets and force the formation of a bow shock within the outer lobes. On the other hand, for the case of a proton-electron plasma we find $\beta_{\rm s} = 0.3$. The propagation of the bow shocks through the outer lobes is very fast. The jets are therefore almost ballistic as predicted by the numerical simulations. 

We can also directly study the momentum balance between the jet and the receding material in the outer lobe. In the rest-frame of the bow shock, momentum balance requires (Bicknell 1994)

\begin{equation}
\left( \gamma_{\rm sj}^2 \beta_{\rm sj}^2 w_{\rm j} + p_j \right) A_j = n_i \bar{m} \beta_{\rm s}^2 c^2 A_{\rm s},
\end{equation}

where $w_{\rm j}$ and $p_{\rm j}$ are the enthalpy and the pressure in the jet with a cross-section $A_j$. For a jet dominated by relativistic particles $w_{\rm j} \sim 4 p_{\rm j}$. $A_{\rm s}$ is the area of the working surface of the bow shock, while $\beta_{\rm sj}$ is the velocity and $\gamma_{\rm sj}$ the Lorentz factor of the jet material measured in the rest-frame of the bow shock. It can be shown that the relativistic addition of velocities results in 

\begin{equation}
\gamma_{\rm sj}^2 \beta_{\rm sj}^2 = \gamma_{\rm s}^2 \beta_{\rm j}^2 \left( \beta_{\rm j} - \beta _{\rm s} \right)^2,
\end{equation}
where $\beta _{\rm j}$ and $\gamma_{\rm j}$ are the velocity and Lorentz factor of the jet material in the rest-frame of the unshocked lobe material. Therefore

\begin{equation}
\left[ 4 \gamma_{\rm s}^2 \gamma_{\rm j}^2 \left( \beta_{\rm j} - \beta_{\rm s} \right)^2 + 1 \right] p_{\rm j} A_{\rm j} = n_i \bar{m} \beta_{\rm s}^2 c^2 A_{\rm s}.
\end{equation}

The jet power of a jet dominated by relativistic particles, i.e. the ratio of rest-mass energy density to internal energy of the jet material, ${\cal R}$, approaches zero, is (Bicknell 1995)

\begin{equation}
Q = 4 c A_{\rm j} p_{\rm j} \gamma_{\rm j}^2 \beta_{\rm j}.
\end{equation}

Solving for $p_{\rm j} A_{\rm j}$ and substituting into the equation for momentum balance we find

\begin{equation}
A_{\rm s} = \frac{4 \gamma_{\rm s}^2 \gamma_{\rm j}^2 \left( \beta_{\rm j} - \beta_{\rm s} \right)^2 +1}{4 \beta_{\rm j} \beta_{\rm s}^2 c^3 \gamma_{\rm j}^2 n_i \bar{m}} Q.
\end{equation}

This expression for the area of the working surface of the bow shock is not very sensitive to the velocity of the jet material. For a jet power $Q=10^{38}$\,W, reasonable values of $\gamma _{\rm j} \le 10$ and again assuming a proton-electron plasma in the outer lobes, we get $A_{\rm s} \sim 100$\,kpc$^2$ corresponding to a radius of $R_{\rm s} \sim 6$\,kpc. This is in good agreement with the observed width of the middle lobes at their leading edges. 

We can apply the same arguments to the inner lobes and interpret them as bow shocks inside the outer lobe structure. In principle the inner lobes may in fact be located inside the middle lobes. However, if the middle lobes are indeed bow shocks, then they will not form wide lobes and hence it is more likely that the inner lobes are also located inside the outer, rather than the middle, lobes. In this case, the magnetic field strength inside the inner lobes is 2\,nT and the electron density is 9\,m$^{-3}$. A pure pair plasma can again be ruled out, because it leads to a bow shock advance speed in excess of $c$. For a proton-electron plasma we find $\beta_{\rm s} = 0.5$ and $R_{\rm s} \sim 3$\,kpc. The inner lobes are consistent with both model interpretations. 

For the interpretation of the inner and the middle lobes as jet-driven bow shocks within the outer lobes we require a gas density in the outer lobes equivalent to at least one proton per relativistic electron. This is a lower limit and the real density in the outer lobes may be higher. Clearly the protons may have been transported into the outer lobes by the jets that inflated them. However, this would significantly increase the already large power of these jets estimated in the previous sub-sections on the basis of the radio luminosity of the outer lobes alone. The alternative is that the protons have ``contaminated" the lobes across their surface. Kaiser et al. (2000) found that fluid instabilities on the lobe surface are slow to develop sufficiently to lead to significant mixing across the lobe boundary. Instead, formation of an atmosphere of sufficient density for the ram pressure confinement of the younger jets could take place via the dispersion of warm gas clouds. These clouds are passing from the ambient medium through the bow shock into the lobes where they are slowly disrupted by turbulent motions (e.g. Gregory et al. 1999). As suggested by Kaiser et al. (2000), the gas stripped off warm clouds inside the lobes may provide a dense enough medium to confine the middle lobes of B0925+420.

\section{Comparison with Galactic X-ray transients}

There is great interest, currently, in comparison of AGN and Galactic black hole candidates, in particular those Galactic transients which vary rapidly, typically showing outbursts which last a few months (e.g. Brocksopp, Bandyopadhyay \& Fender 2004; Brocksopp et al. 2006). Correlations between the X-ray and radio emission have been found, extending over many orders of magnitude in both luminosity and mass (Merloni, Heinz \& di Matteo 2003; Falcke, K\"ording \& Markoff 2004). Furthermore there are suggestions that certain types of AGN are analogous to phases of jet/disc activity in the Galactic transients (K\"ording, Jester \& Fender 2006). For the Galactic transients, jet activity is very closely linked with the X-ray spectral behaviour of the accretion disc and corona mass. Repeated interactions with a companion galaxy was postulated as a mechanism by which the accretion rate could similarly be varied (Schoenmakers et al. 2000a). Of course, the timescales for variability in the AGN are just too long for us to make a direct comparison between AGN and Galactic transients. The DDRG provide a valuable link in this picture; while they still vary on unobservable timescales, we are now able to obtain evidence for previous activity and thus obtain a more accurate idea of how the variability of jets in AGN and Galactic transients can be compared.

A further similarity between the Galactic transients and some lobes of some DDRG is the absence of hotspots (e.g. in the outer lobes, and perhaps the middle lobes, of B0925+420 -- see Section 2). Very few of the Galactic objects have had extended emission from lobes or shocks detected, with the exception of e.g. Cyg X-1 (Gallo et al. 2003), 1E1740.7-2942 (Mirabel et al. 1992) and perhaps GRS 1915+105 (Kaiser et al. 2004). Galactic sources and the DDRG have powerful radio jets, the energy of which needs to be dissipated in the ISM/IGM. If this is not through shocks at the head of the jet (i.e. hotspots) then it is not clear exactly how this dissipation takes place. It is particularly difficult to study this in the Galactic sources since it is rare that the jet emission is resolved; observations of both the jet and the bow shock in Cyg X-1 have led to suggestions that the jet energy may be dissipated via the transport of protons (Heinz 2006). For the DDRG, it seems more likely that, over the large timescales involved, the jets have simply stopped providing energy to the lobes and any residual energy can be gradually dispersed in the IGM.

\section{Conclusions}

We present radio images of B0925+420, the first known ``triple-double'' FRII radio galaxy and resolve, for the first time, the inner pair of radio lobes. B0925+420 provides the strongest evidence yet that the class of DDRGs represent multiple episodes of jet activity and are not just co-aligned knots within a fainter jet. We also image this galaxy in linear polarisation and find that the lobes are polarised up to $\sim 20\%$. We model the flux densities and sizes of the lobes to determine permitted ranges of jet power in terms of the age of the source and density of the surrounding medium. The outer and inner lobes are consistent with the model, and we conclude that the inner lobes lie within the cocoon of the original jet. The middle lobes are more complicated, requiring higher densities than those found within the outer lobes. An alternative model, interpreting the middle and inner ``lobes'' as additional bow shocks within the outer lobes, may be more successful. We suggest that some additional gas has refilled the region, perhaps via the dispersion of warm gas clouds.

\section*{acknowledgments}
The National Radio Astronomy Observatory is a facility of the National Science Foundation, operated under cooperative agreement by Associated Universities, Inc. We are very grateful to Amy Mioduszewski for answering many questions about the data-reduction and to the referee, Dhruba Saikia, for a very detailed and helpful report.


\begin{thebibliography}{}
\bibitem[]{}Baars J.W.M., Genzel R., Pauliny-Toth I.I.K., Witzel A., 1977, A\&A, 61, 99
\bibitem[]{}Baldwin J.E., 1982, p21, in Heeschen D.S., Wade C.M., Eds., Extragalactic Radio Sources, Proceedings of Symposium, Albuquerque, NM, 1981, Dordrecht, D. Reidel Publishing Co.
\bibitem[]{}Becker R., White R., Helfand D., 1995, ApJ, 450, 559
\bibitem[]{}Bicknell G.V., 1994, ApJ, 422, 542
\bibitem[]{}Bicknell G.V., 1995, ApJ Supp., 101, 29
\bibitem[]{}Blundell K., Rawlings S., Willott C.J., 1999, AJ, 117, 677
\bibitem[]{}Brocksopp C., Bandyopadhyay R.M., Fender R.P., 2004, NewA, 9, 249
\bibitem[]{}Brocksopp C. et al., 2006, MNRAS, 365, 1203
\bibitem[]{}Clarke D.A., Burns J.O., 1991, ApJ, 369, 308
\bibitem[]{}Condon J.J., Cotton W.D., Greisen E.W., Yin Q.F., Perley R.A., Taylor G.B., Broderick J.J., 1998, AJ, 115, 1693
\bibitem[]{}Falcke H., K\"ording E., Markoff S., 2004, A\&A, 414, 895
\bibitem[]{}Fanaroff B.L., Riley J.M., 1974, MNRAS, 167, 31P
\bibitem[]{}Fukazawa Y., Makishima K., Ohashi T., 2004, PASJ, 56, 965
\bibitem[]{}Gallo E., Fender R.P., Kaiser C.,  Russell D., Morganti R., Oosterloo T., Heinz S., 2005, Nature, 436, 819
\bibitem[]{}Gregory G., Miniati F., Ryu D., Jones T.W., 1999, ApJ, 527, L113
\bibitem[]{}Heinz S., 2006, ApJ, 636, 316
\bibitem[]{}Jeyakumar S., Saikia D.J., 2000, 311, 397
\bibitem[]{}Kaiser C.R., Alexander P., 1997, MNRAS, 286, 215 (KA)
\bibitem[]{}Kaiser C.R., Cotter G., 2002, MNRAS, 336, 649
\bibitem[]{}Kaiser C.R., Dennett-Thorpe J., Alexander P., 1997, MNRAS, 292, 723 (KDA)
\bibitem[]{}Kaiser C.R., Schoenmakers A.P., R\"ottgering H.J.A., 2000, MNRAS, 315, 381
\bibitem[]{}Kaiser C.R., Gunn K.F., Brocksopp C., Sokoloski J.L., 2004, ApJ, 612, 332
\bibitem[]{}Konar C., Saikia D.J., Jamrozy M., Machalski J., 2006, MNRAS, 372, 693
\bibitem[]{}K\"ording E., Fender R.P., Migliari S., 2006, MNRAS, 369, 1451
\bibitem[]{}K\"ording E., Jester S., Fender R., 2006, MNRAS, 372, 1366
\bibitem[]{}Landau L.D., Lifshitz E.M., 1987, Fluid mechanics, 2$^{\rm nd}$ edition, Butterworth-Heinemann, Oxford
\bibitem[]{}Longair M.S., 1994, High energy astrophysics, Cambridge University Press
\bibitem[]{}Merloni A., Heinz S., di Matteo T., 2003., MNRAS, 345, 1057
\bibitem[]{}Mirabel I.F., Rodriguez L.F., Cordier B., Paul J., Lebrun F., 1992, Nature, 358, 215
\bibitem[]{}Rengelink R.B., Tang Y., de Bruyn A.G., Miley G.K., Bremer M.N., R\"ottgering H.J.A., Bremer M.A.R., 1997, A\&AS, 124, 259
\bibitem[]{}Saikia D.J., Konar C., Kulkarni V.K., 2006, MNRAS, 366, 1391
\bibitem[]{}Schoenmakers A.P., de Bruyn A.G., R\"ottgering H.J.A., van der Laan H., Kaiser C.R., 2000a, MNRAS, 315, 371
\bibitem[]{}Schoenmakers A.P., de Bruyn A.G., R\"ottgering H.J.A., van der Laan H., 2000b, MNRAS, 315, 395 
\bibitem[]{}Spergel D.N. et al., 2003, ApJS, 148, 175








\end{thebibliography}
\end{document}